**Non-volatile ferroelastic switching of the Verwey transition and resistivity of epitaxial $Fe_3O_4$/PMN-PT (011)**

Ming Liu,[1*] Jason Hoffman,[2] Jing Wang,[3] Jinxing Zhang,[3*] Brittany Nelson-Cheeseman[2] and Anand Bhattacharya[1,2*]

1. Center for Nanoscale Materials, Argonne National Laboratory, Argonne, IL 60439 (USA)
2. Material Science Division, Argonne National Laboratory, Argonne, IL 60439 (USA)
3. Department of Physics, Beijing Normal University, Beijing 100875 (China)
[*] mingliu@anl.gov, jxzhang@bnu.edu.cn, anand@anl.gov

Keywords: Verwey transition, Magnetoelectric coupling, Multiferroics, Metal-insulator transition, Non-volatile, Mottronics

**Abstract**:

A central goal of electronics based on correlated materials or 'Mottronics' is the ability to switch between distinct collective states with a control voltage. Small changes in structure and charge density near a transition can tip the balance between competing phases, leading to dramatic changes in electronic and magnetic properties. In this work, we demonstrate that an electric field induced two-step ferroelastic switching pathway in (011) oriented $0.71Pb(Mg_{1/3}Nb_{2/3})O_3$-$0.29PbTiO_3$ (PMN-PT) substrates can be used to tune the Verwey metal-insulator transition in epitaxial $Fe_3O_4$ films in a stable and reversible manner. We also observe robust non-volatile resistance switching in $Fe_3O_4$ up to room temperature, driven by ferroelastic strain. These results provides a framework for realizing non-volatile and reversible tuning of order parameters coupled to lattice-strain in epitaxial oxide heterostructures over a broad range of temperatures, with potential device applications.

The central challenge in realizing electronics based on strongly correlated materials, or 'Mottronics',[1] is the ability to switch between distinct electronic and magnetic phases with a control voltage. These phases include correlated insulators and various magnetic states that result from strong electron-electron and electron-lattice interactions. Small changes in structure and charge density near a transition between competing phases can tip the balance among them, leading to large changes in electronic and magnetic properties. Devices based upon such transitions could be, in principle, both fast and energy efficient,[2] overcome some of the intrinsic limitations in conventional field-effect transistors[3] and also provide new functionalities.[4-6] Approaches based upon the electrostatic field-effect are thought to require gate-induced charge densities in excess of $1 \times 10^{14}$ carriers/cm$^2$, too high for conventional dielectrics.[4] In recent years, this challenge has been addressed using ferroelectric[7] and ionic liquid dielectrics,[8,9] though in the latter electrochemical effects can play a very important role.[10] In principle, electrostatic gating only affects a region comparable to the Thomas Fermi/Debye screening length near the interface of the material with the gate dielectric. On the other hand, epitaxial strain imposed by a substrate can be used to tune the properties of a thin film through its entire thickness. Notably, this has been used to realize novel multiferroicity in EuTiO$_3$[11,12], manipulate the metal-insulator transition in perovskite manganites[13-15], change ferroelectric and ferromagnetic transition temperatures[16-21] and enable flexoelectric rotation of polarization in PbTiO$_3$.[22] In these systems in-plane epitaxial strain is used to control lattice-spin, lattice-phonon, and lattice-charge interactions to tailor properties or realize new magnetic/electronic phases. While these experiments point to a pathway for creating novel functionalities, reversible and non-volatile switching between different phases using strain has remained largely unexplored. In non-volatile switching, the electronic/magnetic states that are realized remain in a stable remnant state after the control voltage has been switched off. Furthermore, for memory applications, we need to be able to switch between these different states in a reversible manner. One of the most promising approaches for *in-situ* manipulation of lattice-coupled order parameters is to grow oxide films on ferroelectric/piezoelectric substrates.[21,23-27] Application of electric fields in these structures induces changes in the substrate lattice parameters, resulting in changes in the properties of the films. This has been demonstrated, for example, in $La_{0.7}Sr_{0.3}MnO_3/BaTiO_3$,[21] $La_{0.7}Sr_{0.3}CoO_3$/PMN-PT(001),[28] and $La_{0.7}Ca_{0.15}Sr_{0.15}MnO_3$/PMN-PT(001) heterostructures[29] (here, PMN-PT is an alloy of $Pb(Mg_{1/3}Nb_{2/3})O_3$ (71%) and $PbTiO_3$ (29%), and has a very large piezo response). However, in these studies non-volatile and reversible switching could not be demonstrated since the strain resulting from a piezo response decays upon removal of the control voltage, and in a purely ferroelectric reversal the strain states in the two stable polarization directions are equivalent. In our work, we use a unique *ferroelastic*[30] switching pathway in (011) oriented PMN-PT, that allows polarization vectors to rotate from an out-of-plane to a purely in-plane direction to produce two distinct, stable, non-volatile and reversible lattice strain states.[31] Using these two strain states, we demonstrate electric-field control of the Verwey metal-insulator transition, and non-volatile reversible resistive switching in epitaxial $Fe_3O_4$/PMN-PT (011) heterostructures over a broad range of temperatures.

Magnetite is a conducting spin-polarized ferrimagnet with resistivity > 1 mΩ-cm at 300 K.[32] It undergoes a Verwey metal-insulator transition as the temperature drops below $T_V$ = 125 K, with an abrupt increase in resistivity by 2-3 orders of magnitude and a slight decrease in magnetic moment.[32] This transition is associated with a symmetry lowering structural transition from cubic inverse-spinel to monoclinic, and is therefore greatly sensitive to lattice or structural changes. The mechanism behind the Verwey transition is the subject of a long-standing debate between charge ordering ('Wigner crystal') of the $Fe^{2+}/Fe^{3+}$ cations on the octahedral sites and structural/orbital ordering, and remains an active area of research.[33] The transition may be suppressed with hydrostatic pressure in bulk single crystalline magnetite,[34] where metallization[35] and subtle structural and spin-state transitions[36] have been observed at low temperatures and pressures above 6 GPa. However this approach requires a large volume change induced by mechanical force, which may not be suited for device applications. In this work, we grow epitaxial $Fe_3O_4$ films on (011) oriented PMN-PT substrates. Through ferroelastic control of lattice strain, we suppressed the Verwey metal-insulator transition by up to 8 K, evidenced by large changes in resistivity, magnetization and magnetoresistance. Further, we address a critical requirement for device applications with the discovery of a reversible and non-volatile ferroelastic strain induced resistance switching in $Fe_3O_4$ that persists up to room temperature. Our approach offers new possibilities for novel electronic devices, frequency-agile microwave applications and practical realizations of non-volatile memories.[37-42]

**Results**

**Ferroelastic control of the Verwey transition in epitaxial $Fe_3O_4$/PMN-PT (011)**. A key aspect of our study is the ability to grow epitaxial $Fe_3O_4$ films with a sharp Verwey transition on (011) oriented PMN-PT substrates by ozone-assisted molecular-beam epitaxy. The growth conditions were refined by depositing epitaxial $Fe_3O_4$ films on MgO (001), $MgAl_2O_4$ (001), PMN-PT (001), and PMN-PT (011) substrates and carrying out detailed structural, magnetic and electrical characterization (Fig S1-S4). The thicknesses of all films were determined to be around 55 nm by fitting to x-ray reflectivity spectra. A well-defined Verwey transition was observed in $Fe_3O_4$ films on all substrates as determined by resistivity and magnetization measurements (Fig. 1, Figs. S3,S4).

In order to explore ferroelastic tuning of $Fe_3O_4$ on PMN-PT (011), the electronic transport properties of magnetite were first studied under two distinct electric-field induced polarization states of the substrate. They are the initial or unpoled state denoted by $P_0$, and the poled remnant state denoted by $P_r$. Figure 1(a) shows the schematic of the experimental setup. The film resistance was measured by a four-probe technique. Poling PMN-PT(011) was achieved by applying an electric field in the (011) direction, using the $Fe_3O_4$ film as the top electrode and a sputtered gold film as the bottom electrode. The top electrode was always held at ground when the poling voltage was applied.

We carried out *in situ* x-ray diffraction measurements on $Fe_3O_4$/PMN-PT (011) heterostructures in the unpoled and poled strain states as shown in Fig. 1(b). An expansion along the out-of-plane direction associated with an effective in-plane contraction was observed as the $Fe_3O_4$/PMN-PT (011) was poled. The enhanced out-of-plane lattice parameters of PMN-PT(011) and $Fe_3O_4$ are visible as a shift of $\Delta c/c = + 0.26$ % and $+ 0.20$% respectively. This difference results presumably from different Poisson's ratios of the two materials. The strain state dependence of lattice parameter in $Fe_3O_4$/PMN-PT (011) indicates that an electric-field-induced ferroelastic strain can be coherently transferred to the film, influencing magnetic and transport properties as discussed below.

Figure 1(c) shows the temperature dependence of the resistivity in $Fe_3O_4$/PMN-PT (011) under unpoled $P_0$ and poled $P_r$ states (note logarithmic scale for resistivity). At $P_0$ state (blue curve), the resistivity of the film increases with reducing temperature and undergoes a metal-insulator transition at $T_v$ = 125 K. Once the sample is poled with a DC field of 10 kV/cm, the resistivity decreases dramatically over a broad temperature range, with a visible reduction in the Verwey temperature of ~ 8 K. The large resistance change is more clearly visible in the inset of Fig. 2(c) (linear scale for resistivity). Our results reveal that an in-plane compressive strain in the poled $P_r$ state inhibits the Verwey transition and decreases $T_V$. Using the measured $\Delta c/c$ of 0.2% and Poisson's ratio for $Fe_3O_4$, we calculated the volume change produced by in-plane strain in $Fe_3O_4$ films to be 0.3%~0.4% (Fig. S5). In previous studies[34] of the effects of hydrostatic pressure on bulk magnetite, an order of magnitude higher volume change of ~ 3% was required to produce an equivalent suppression of $T_V$ by ~ 8 K. Thus, electrically modulating the in-plane epitaxial strain is far more effective in manipulating the Verwey transition than would be expected from a pure volume change argument (Calculation details in Supplementary).

Figure 1(d) shows the relative resistance change between the poled and unpoled polarization states, defined as $\Delta R = (R_0 - R)/R_0$. Upon decreasing temperature, $\Delta R$ increases continuously from ~ $10^2$ Ω at 250 K to ~$10^5$ Ω at 90 K (three orders of magnitude) in Fe$_3$O$_4$/PMN-PT (011). However, $\Delta R/R$ has a maximum value of 88% at 125 K, close to $T_V$, indicating that the resistivity is maximally sensitive to structural or lattice change at the Verwey transition. In addition, $\Delta R/R$ were found to be 3-5% and 30% at 300 K and 90 K, respectively. This result is in contrast to a recent report, where the resistance of a magnetite film *increased* near the Verwey transition by electrostatic field-effect gating in an ambipolar manner (i.e., the resistance increased for both induced holes, as well as electrons).[43] This indicates that strain effects and not electrostatic charging play the dominant role in our structures, where the strain state and not the polarity of **P** determines the sign of resistance change.

The polarization states in PMN-PT(011) not only affects the electronic structure but also result in magnetization change in magnetite around $T_V$. As shown in Figure 1(e), the abrupt change in magnetization was suppressed to lower temperature, consistent with a decrease in $T_V$. The decrease was quantitatively determined to be ~ 8 K from the first derivative of magnetization versus temperature (inset), consistent with the resistivity measurements in Fig. 1(c). Figure 1(f) shows the strain state dependence of *M-H* loops at $T_V =$ 125 K, exhibiting distinct changes in magnetic anisotropy and magnetization. Such effects also arise from the suppression of the structural transition at $T_V$ by electrically poling the PMN-PT (011), which leads to a change in the magneto-crystalline anisotropy as well as magnetization. Thus, we have successfully realized ferroelastic tuning of the Verwey metal-insulator transition as well as associated electronic and magnetic states.

**Electrical control of non-volatile and reversible resistivity switching**. Due to the stability of remnant in-plane polarization in (011) oriented PMN-PT, stable and reversible ferroelastic switching, as well as non-volatile resistance tuning can be realized in Fe$_3$O$_4$/PMN-PT (011).

Details of ferroelastic switching of PMN-PT (011) are discussed later. Figure 2(a) shows *in-situ* control of resistance using an applied voltage in $Fe_3O_4$/PMN-PT (011) at various temperatures. Hysteretic changes in resistance were observed upon applying appropriate unipolar electric fields across the PMN-PT substrate, showing a large tunability of up to 50% change in resistance near $T_V$. The two remnant resistance states in the hysteresis loops represent two ferroelastic strain states in PMN-PT(011), which are stable and switchable. We note that the non-volatile change in resistance here is 50%, which is smaller than that observed in Fig. 1(c). We attribute this to an increase in the coercive field of PMN-PT(011) at low T, resulting in incomplete polarization switching upon poling at these temperatures.[44] Electrical modulation of resistance can also be achieved at room temperature, as shown in Fig. 2(b). With appropriate application of unipolar and bipolar electric fields, both a resistance hysteresis loop (blue) and 'butterfly' curve (red) were observed. Figure 2(c) shows the first derivatives of *ln*(R) with respect to the applied electric field (E) across the PMN-PT(011) substrate at various temperatures. Resistance change coefficients up to 35 %-cm/kV at room temperature and 28 %-cm/kV at 125 K were observed as the polarization vector was electrically switched between in-plane and out-of-plane near the coercive fields. The sharpness of the peaks in |d(ln(R))/dE| and the lower coercive fields at room temperature make these more suitable for device applications. Conversely, at low T the broad peaks and high coercive fields point to relatively difficult ferroelastic polarization switching of the PMN-PT.

Figure 2(d) shows electric-field-induced non-volatile resistance switching in $Fe_3O_4$/PMN-PT(011) at room temperature and $T_V$ ~ 125 K. The resistance was modulated when an appropriate square wave of electric field was applied across the PMN-PT (blue) to switch between out-of-plane and in-plane polarization of the PMN-PT, giving rise to two distinct non-volatile and stable strains states. As a consequence, we were able to tune the resistance by 3% and 50%, at room temperature and $T_V$ respectively. The resistance response follows the excitation of the electric field and did not show significant decrease after being cycled for 5 hours with a cycling frequency of 0.25 Hz.

**Ferroelastic switching pathway in (011) oriented PMN-PT**. To understand the evolution of ferroelastic domains during switching in (011) oriented PMN-PT, we studied these substrates using piezoresponse force microscopy (PFM). We imaged the polarization domains using both vertical (V) and lateral (L) PFM measurements.[45] In VPFM, we are sensitive to the vertical piezo response, which is used to image the out-of-plane polarization component. In LPFM we measure torsional displacements that are sensitive to the in-plane component perpendicular to the cantilever, which is aligned parallel to the scanning direction (Fig. 3(a)). Thus, for a scan direction along <100>, we are sensitive to the in-plane polarization projection along the <0-11> directions. As is evident from the schematic (top row of Fig. 3), this component of polarization is non-zero only when the polarization lies in the (011) plane (parallel to the sample surface). Figure 3(a)-3(c) show schematics of polarization vectors, and domain images of the unpoled PMN-PT (011) crystal substrate. In the unpoled state at room temperature, the polarization vectors randomly point along the 8 body diagonals of the pseudocubic cell with rhombohedral symmetry, including 4 equivalent in-plane directions. This gives rise to the PFM phase images shown in Fig. 3(b) and 3 (c), showing a mixture of ferroelastic domains. When a poling voltage of +10V is applied on the tip, both ferroelectric switching ($180^0$) and ferroelastic switching ($71^0$ and $109^0$) take place and all the polar vectors rotate downward in the poled area (blue box), pointing along two equivalent directions, as shown in Fig. 3(d). In this orientation, the polarization has no component along the <0-11> directions and thus there is no contrast within the blue box of the LPFM image in Fig. 3(f). Further, the average in-plane strain is compressive compared to the unpoled state, explaining the shift in lattice parameters shown in Fig. 1(b). As a subsequent DC voltage of -5V was applied, the polarization vectors change to an intermediate state with a large majority of the sample polarization lying purely in the (011) plane without any out-of-plane component, as shown in Fig. 3(g). In this orientation, domains of polarization pointing along <0-11> are clearly evident within the red box of the LPFM image in Fig. 3 (i). When the voltage was further increased to -8V, all of the out-of-plane polarization reversed as shown in Fig. 3(k) and the pure in-plane polarization completely disappeared (Fig. 3(l)). Further details of switching dynamics can be seen in the supplementary section (Figs. S6-S9). These results indicate that upon increasing the poling bias gradually, a two-step 71° or 109° ferroelastic switching takes place, which first makes the polarization components completely rotate from the downward direction into (011) in-plane directions, and subsequently point in the upward direction at the end of the reversal. This is suppressed in thin film ferroelastic systems due to epitaxial clamping by the substrate. The stable and reversible in-plane and out-of-plane polarization states produced here enables non-volatile resistance tuning in $Fe_3O_4$/PMN-PT (011) upon applying appropriate electric fields. In this process, strain or lattice change in epitaxial $Fe_3O_4$ films arises due to ferroelastic switching in PMN-PT (011), which was determined to cover 70% of the whole poled area (Details in supplementary in Fig. S9). This highlights the extraordinary nature of poling PMN-PT (011) in comparison to PMN-PT (001),[46] with dominant ferroelastic switching in PMN-PT (011) occurring due to the existence of a stable in-plane polarization state.

**Role of anti-phase boundaries in ferroelastic tuning of $Fe_3O_4$/PMN-PT (011).** In epitaxial magnetite films, atomically sharp anti-phase boundaries (APBs) form as natural growth defects. Proliferation of a large number of these can strongly suppress the Verwey transition, and it is important to establish their role in ferroelastic switching of the resistance of our samples. This is particularly relevant in light of the large changes in the nanoscale ferroelastic domain patterns that we observe during switching. At the APBs, the exchange between neighboring spins is antiferromagnetic, though the regions on either side are ferromagnetically aligned. From considerations of double-exchange mediated transport and canting of spins, it can be shown that this leads to a large negative and linear magnetoresitance (MR) at high magnetic fields.[47,48] Thus, one way to probe the APBs is to measure the MR in epitaxial Fe$_3$O$_4$ films in different ferroelastic strain states. Figure 4 shows the ferroelastic strain dependence of MR, defined as MR=(R(H)-R(0))/R(0) at various temperatures with applied magnetic fields and currents along the in-plane [100] direction. In the unpoled sample, the negative MR increases in magnitude as T decreases and peaks near $T_V$ (Fig. 4(a) (blue)). Upon poling the substrate, the maximum MR peak shifts to lower temperature by up to 8 K (red), indicating the suppression of $T_V$. However, we note that at higher temperatures (>180 K), there is no measurable change in the MR between the poled and unpoled states. This indicates that there were no significant changes in either the number of APBs or in their magnetic anisotropy due to ferroelastic (compressive) strain. Furthermore, at low temperature, a *decrease* in the magnitude of MR by 1-3% was observed at high magnetic fields for the poled sample (Fig. 4(b)). We relate this phenomenon to ferroelastic strain induced increase in the magnetic anisotropy of Fe spins in the vicinity of the APBs, which would also *increase* the resistance[43] of the APBs. In summary, the large decrease in resistance by 88% upon poling near $T_V$ cannot be attributed to a change in the number of APBs upon poling. Changes in resistance of the APBs due to changes in magnetic anisotropy are not large enough, and also have the wrong sign, to explain the decrease in resistance upon poling.

**Discussion**

In conclusion, we have demonstrated electrical tuning of the Verwey metal-insulator transition and non-volatile resistance switching at room temperature in epitaxial Fe$_3$O$_4$/PMN-PT (011) heterostructures. This was achieved using a unique ferroelastic switching pathway in PMN-PT (011) to maximize ferroelastic strain and realize two distinct, stable and reversible strain states. These results point to opportunities for electrical tuning of strain sensitive phase transitions and other properties of correlated oxides that may be grown on PMN-PT(011) substrates.

**Methods**

The $Fe_3O_4$ films were epitaxially synthesized on (011) oriented PMN-PT single crystal substrates in a MBE chamber with a base pressure in the $10^{-10}$ Torr range. A pure flow of distilled ozone was delivered to the chamber as the oxidizing agent, maintaining the chamber pressure at $1.0 \times 10^{-7}$ Torr at the growth temperature of 300 °C. 99.9% pure Fe was deposited from a differentially pumped dual-filament Knudsen cell. Prior to film growth, the deposition rate was measured using a quartz-crystal thickness monitor (QCM), which is 0.035~0.045 Å/sec with a drift of ~ 0.5% per hour. When deposition was complete, the sample was cooled to room temperature in ozone at the same pressure. This growth conditions were refined by depositing epitaxial $Fe_3O_4$ films on MgO (001), $MgAl_2O_4$ (001) and PMN-PT (001) substrates and carrying out detailed structural, magnetic and electrical characterization (Supplementary).

**Acknowledgements**

Work at Argonne National Laboratory, including use of facilities at the Center for Nanoscale Materials, was supported by the U.S. Department of Energy, Office of Basic Energy Sciences under contract No. DE-AC02-06CH11357. Dr. Ming Liu was supported by a Directors' Postdoctoral Fellowship at Argonne. The work in Beijing Normal University was supported by National Science Foundation of China under contract No.11274045.

**Author contributions**

M.L. and A.B. planned the experiments. M.L. made the samples and performed all magnetic and electronic measurements, with assistance from J.H. and B.N.-C. PFM scans were performed by J.W. and J.Z. The paper was written by M.L and A.B. All authors discussed the results and commented on the manuscript.

**Addition information**

Competing financial interests: The authors declare no competing financial interests.

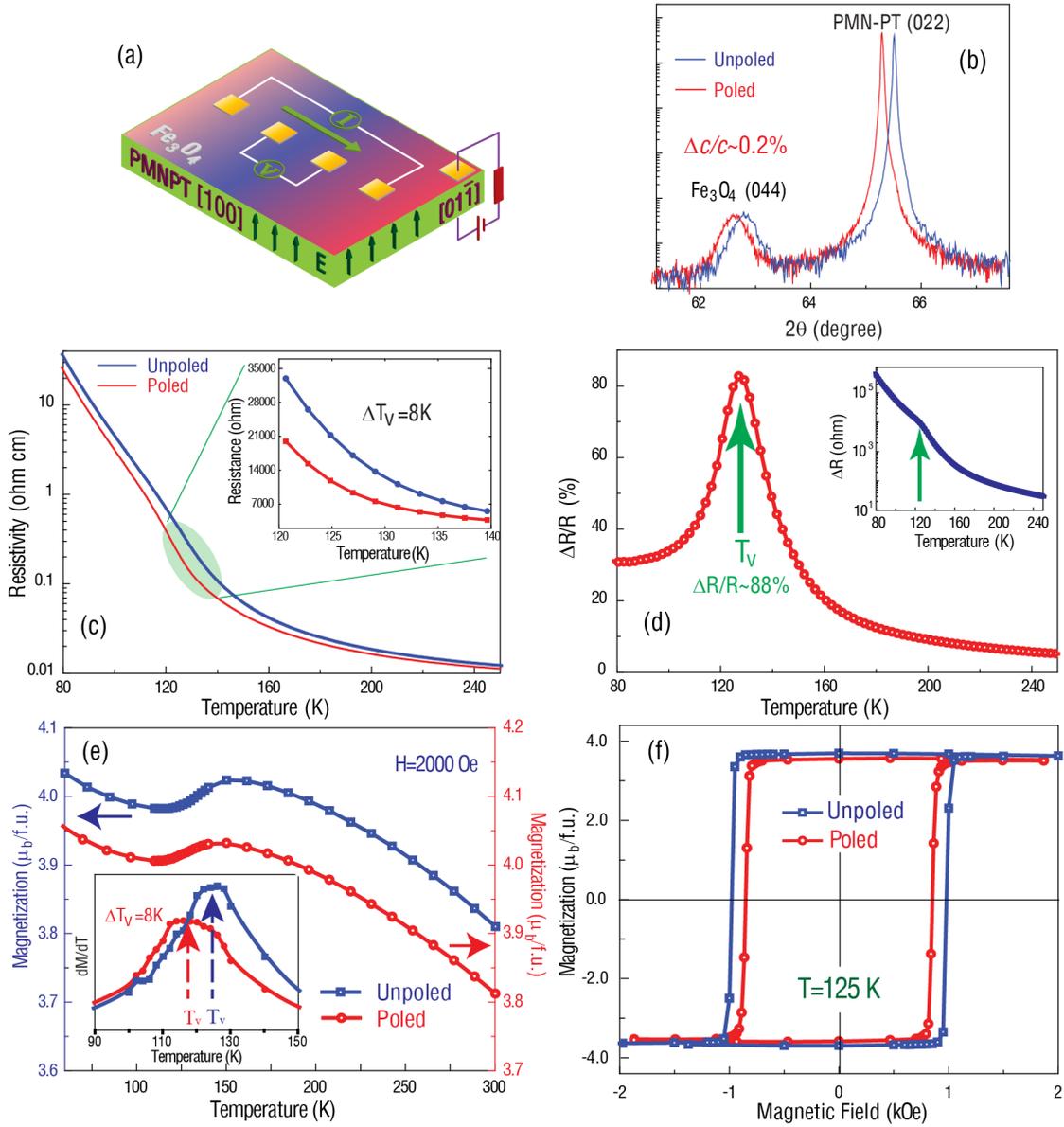

**Figure 1**. **Ferroelastic strain control of Verwey metal-insulator transition in $Fe_3O_4$/PMN-PT (011)**. (a) Schematics of four-point resistance measurement. (b) X-ray diffraction patterns of $Fe_3O_4$/PMN-PT (011) under unpoled and poled states. (c) Ferroelastic strain state dependence of resistivity as a function of temperature for (011) oriented $Fe_3O_4$/PMN-PT. (d) Relative resistance change at a function of temperature. The insets are absolute value of resistance change. (e) Magnetic moment as a function of temperature for poled and unpoled $Fe_3O_4$/PMN-PT(011). Inset shows the first derivative. (f) Magnetic hysteresis loops at Verwey temperature.

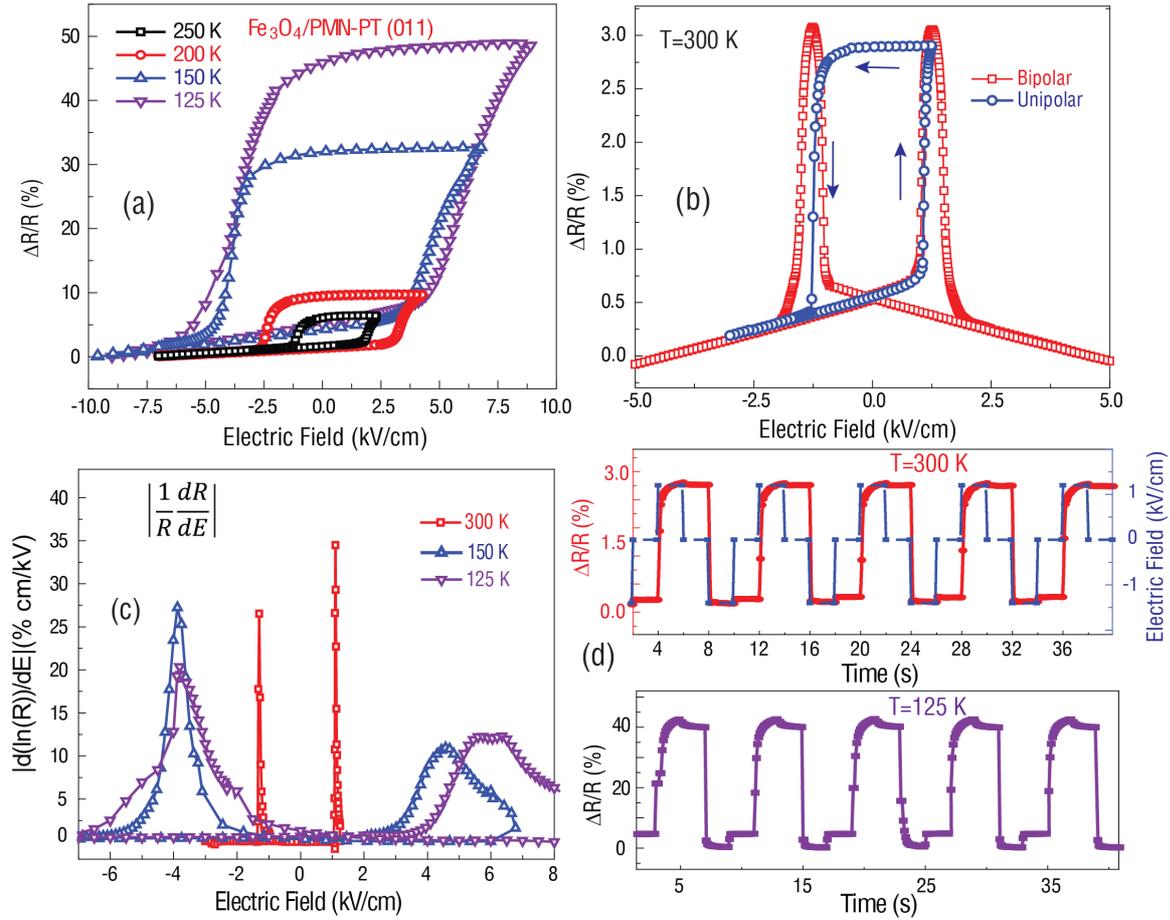

**Figure 2**. **Non-volatile ferroelastic switching of resistivity in in Fe$_3$O$_4$/PMN-PT (011)**. (a) Resistance hysteresis loops at various temperatures. (b) Resistance response of Fe$_3$O$_4$/PMN-PT(011) under unipolar and bipolar sweeping of electric fields at room temperature. (c) The absolute value of the first derivative of the unipolar hysteretic resistance loop with electric field, as a function of electric field. (d) Electric-field-induced non-volatile resistance switching at T=300 K and T=125 K.

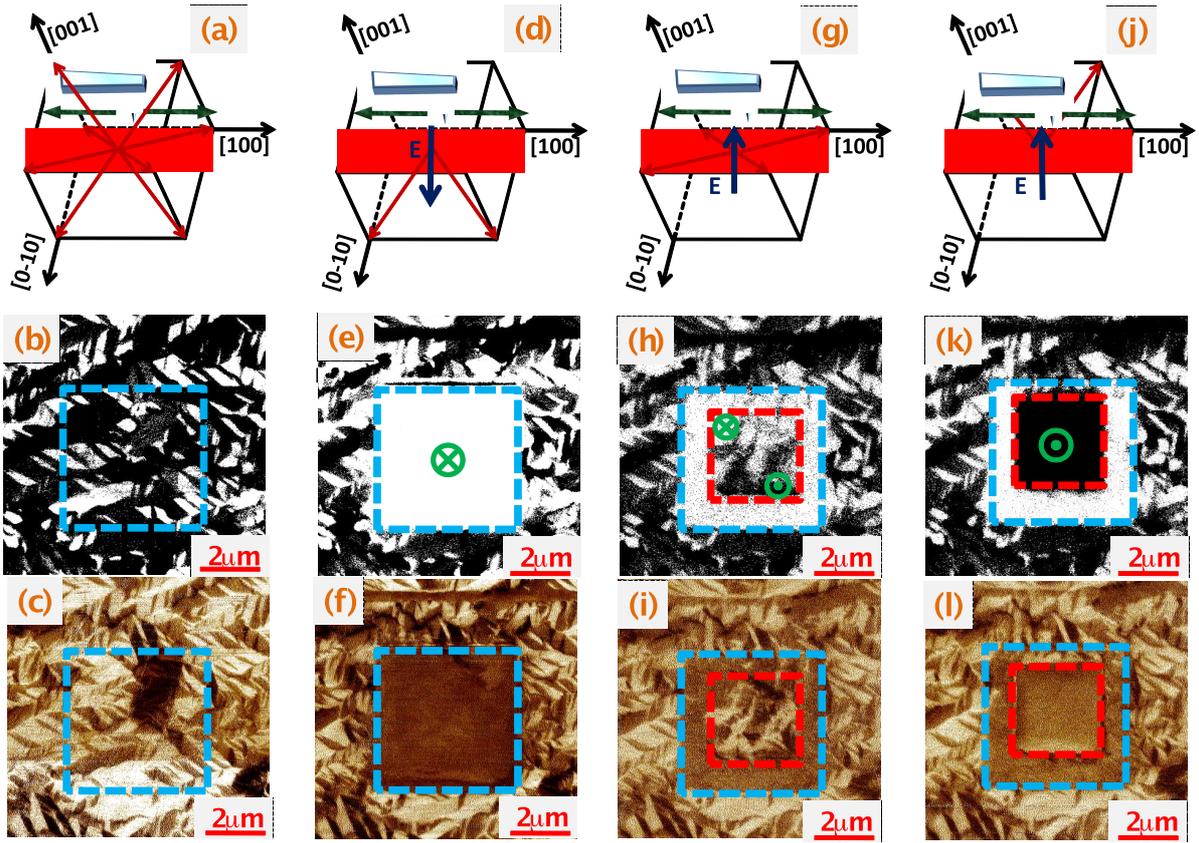

**Figure 3. Probe-bias-dependent nanoscale switching dynamics in the (011) oriented PMN-PT crystal.** (a), (d), (g) and (j) are the schematics showing the scanning direction of the cantilever and the initial, poled, intermediate and final states of the polar vectors. The red arrows along <111> represent the spontaneous polarization vectors in the rhombohedral phase. The green arrows along <100> represent the direction of scanning of the PFM cantilever and the blue arrows in the center of cubes stand for the direction of the electric field. Panels (b), (e), (h) and (k) are the out-of-plane VPFM phase images of the above different switching states, while (c), (f), (i) and (l) are the in-plane LPFM phase images at different poling biases, respectively. These images show a switching process with the increase of tip bias, demonstrating the existence of a pure in-plane polarization at an intermediate voltage.

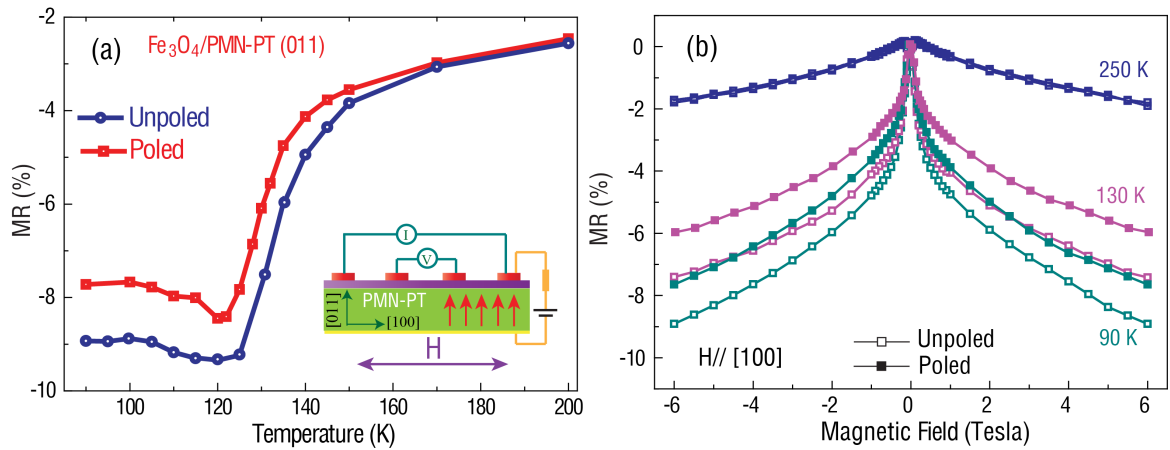

**Figure 4. The effect of ferroelastic strain on magnetoresistance in Fe$_3$O$_4$/PMN-PT (011).** (a) The temperature dependence of magnetoresistance measured at H=6 Tesla under poled and unpoled polarization states. (b) Ferroelastic strain dependence of magnetoresistance at various temperatures with applied magnetic fields and measured current along the in-plane [100] direction.